\documentclass[3p,times,procedia]{elsarticle}
\usepackage{nupha_ecrc}

\volume{00}

\firstpage{1}

\journalname{Nuclear Physics A}

\runauth{}

\jid{nupha}

\jnltitlelogo{Nuclear Physics A}

\usepackage{amssymb}

\usepackage{lineno}

\usepackage[figuresright]{rotating}

\begin{document}

\begin{frontmatter}



\dochead{XXVIIIth International Conference on Ultrarelativistic Nucleus-Nucleus Collisions\\ (Quark Matter 2019)}

\title{Constraining parton energy loss via angular and momentum based differential jet measurements at STAR}


\author{Raghav Kunnawalkam Elayavalli for the STAR Collaboration}

\address{Wayne State University, Detroit MI 48201}

\begin{abstract}
Parton energy loss has been established as an essential signature of the Quark-Gluon Plasma (QGP) in heavy ion collisions since the earliest measurements at RHIC indicating suppression of hadron spectra at high $p_{\rm{T}}$ and coincidence yields. Understanding this phenomenon of jet quenching is a requirement for extracting the microscopic properties of the QGP via jet-tomography. STAR has recently introduced a technique called Jet Geometry Engineering (JGE) wherein we enforce particular selection criteria imposed on the jet collection, such as recoiling off a high $p_{\rm{T}}$ hadron trigger along with an additional transverse momentum threshold for jet constituents in events with back-to-back di-jets. With JGE, we are able to control the extent of energy loss ranging from quenched/imbalanced to recovered/balanced di-jets. Since jet quenching is also expected to be dependent on the resolution/transverse-length scales with which the jet probes the medium, it is necessary to perform differential measurements with a handle on both momentum and angular scales. To quantify the angular scale within jets, we present the first measurement of the jet's inherent angular structure in Au$+$Au collisions at $\sqrt{s_{\mathrm{NN}}} = $ 200 GeV via the opening angle between the two leading sub-jets ($\theta_{\rm{SJ}}$). We also measure the di-jet asymmetry $A_{\rm{J}}$ differentially as a function of the $\theta_{\rm{SJ}}$ observable for these di-jets and find no significant dependence of the energy loss on the opening angle of the recoil jet.
\end{abstract}


\end{frontmatter}


\section{Introduction}
Recent measurements at both RHIC and the LHC along with theoretical advances have shown the importance of studying and measuring the properties of jets produced in both $p$+$p$ and heavy ion collisions (review of jet studies can be found here~\cite{expreview}). Since these jets are defined via clustering algorithms on collections of objects (particles or tracks/towers experimentally), varying the jet finding parameters including the jet radius and constituent momentum thresholds can result in selecting different collections of jets and in the case of di-jets, different types of events. A primary signature of the QGP produced in heavy ion collisions is the phenomenon of jet quenching wherein a hard scattered parton loses energy as it traverses the hot and dense medium. Jet quenching not only reduces the jet's energy, but also results in modification of its internal structure. There are two natural scales that characterize a jet and its evolution: the momentum and the angular scales. With an expanding QGP, it is possible that jet-medium interactions could be dependent on the resolution scale or the coherence length of the medium that perceives the jet as a singular radiating object or a multi-prong object~\cite{cohlength}. In order to disentangle the different effects of jet quenching and to quantitatively extract medium properties, we undertake two different types of differential di-jet measurements of jet quenching.   

\section{Datasets and Analysis details}
The Au$+$Au collision data used in these proceedings were collected during the 2007 run with its corresponding $p$+$p$ reference collected in 2006 at $\sqrt{s_{\mathrm{NN}}} = 200$ GeV by the STAR experiment~\cite{star}. We employ a high tower (HT) trigger, requiring at least one Barrel Electromagnetic Calorimeter (BEMC) tower with $E_{\rm{T}} > $ 5.4 GeV. Event centrality in Au$+$Au events is determined by the raw charged track multiplicity in the Time Projection Chamber (TPC) within $|\eta| < 0.5$ and we show only events in the 0-20\% centrality range. In Au$+$Au collisions, we have two separate jet collections, both clustered with the anti-$k_{\rm{T}}$ algorithm~\cite{FastJet}, given by the HardCore jet selection, where jets are clustered with objects (tracks/towers) with $p_{\rm{T}}> 2$ GeV/$c$, and Matched jet selection, where jets are clustered with $p_{\rm{T}} > 0.2$ GeV/$c$ for constituents and geometrically matched to the HardCore jets~\cite{starprl}. Matched jets are background subtracted to suppress the underlying event contribution via both the area subtraction method, utilized for the di-jet asymmetry, and the constituent subtraction method~\cite{cs} for the substructure studies. Further event selection criteria include a minimum jet $p_{\rm{T}}$ requirement for HardCore di-jets ($p^{\rm{Lead}}_{\rm{T, jet}} > 16, p^{\rm{SubLead}}_{\rm{T, jet}} > 8$ GeV/$c$) and an azimuthal angle ($|\Delta \phi (\rm{Lead, SubLead})| > 2\pi/3$) selection to focus on back-to-back di-jets. 

For a meaningful comparison between results in Au$+$Au collisions and in the $p$+$p$ reference, the effects of background fluctuations and detector inefficiencies must be taken into account. To achieve this, HT-triggered $p$+$p$ collisions are embedded into minimum bias Au$+$Au collisions ($p$+$p$ $\oplus$ Au$+$Au), in the same centrality range (0-20\%). During embedding, we account for the relative tracking efficiency (90\%$\pm$7\%) and relative tower energy scale (100\%$\pm$2\%), with a one sigma variation taken as systematic uncertainties.

\section{Jet Geometry Engineering}

\begin{figure}[h] 
   \centering
   \includegraphics[width=0.45\textwidth]{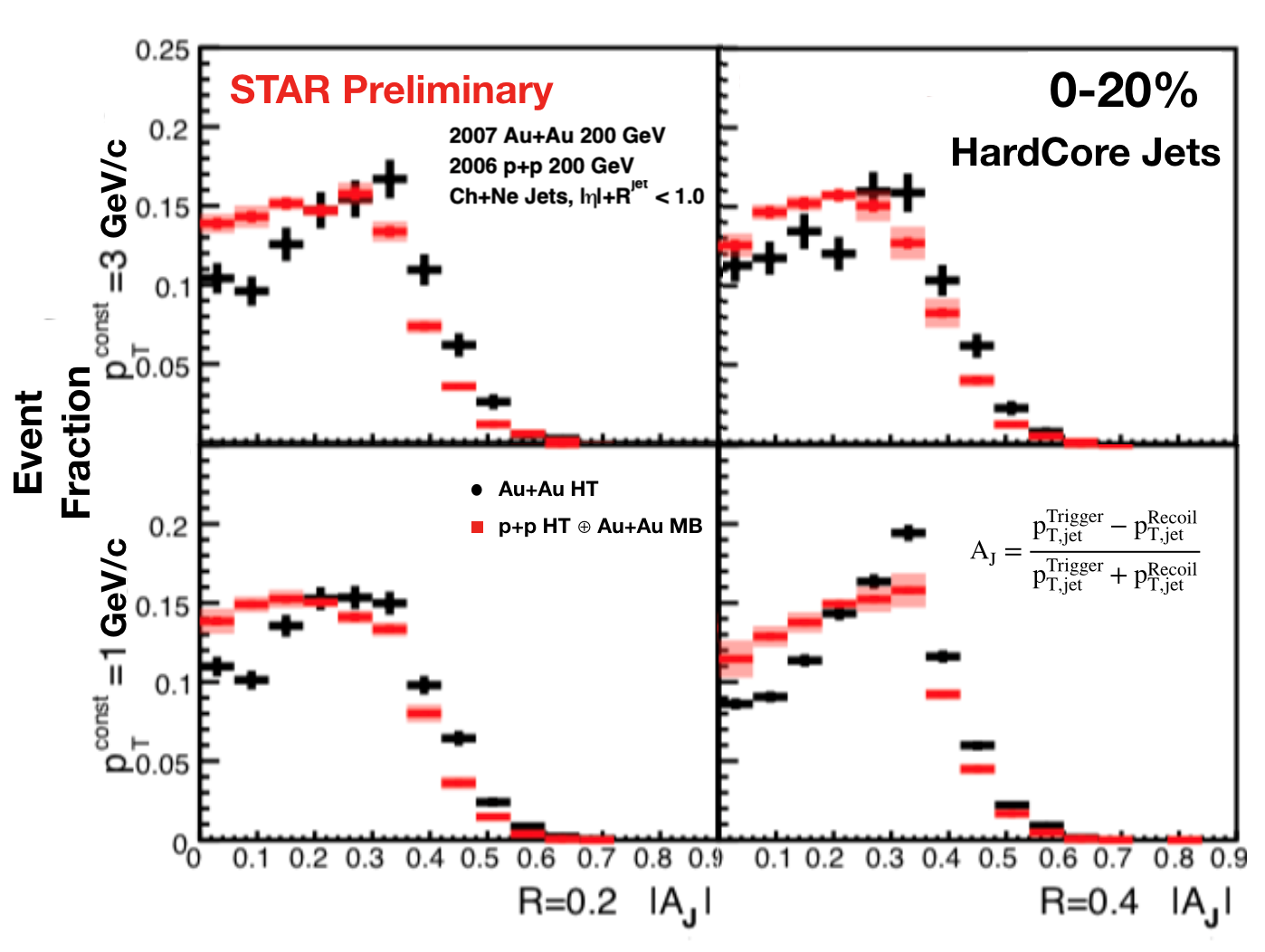}
   \includegraphics[width=0.465\textwidth]{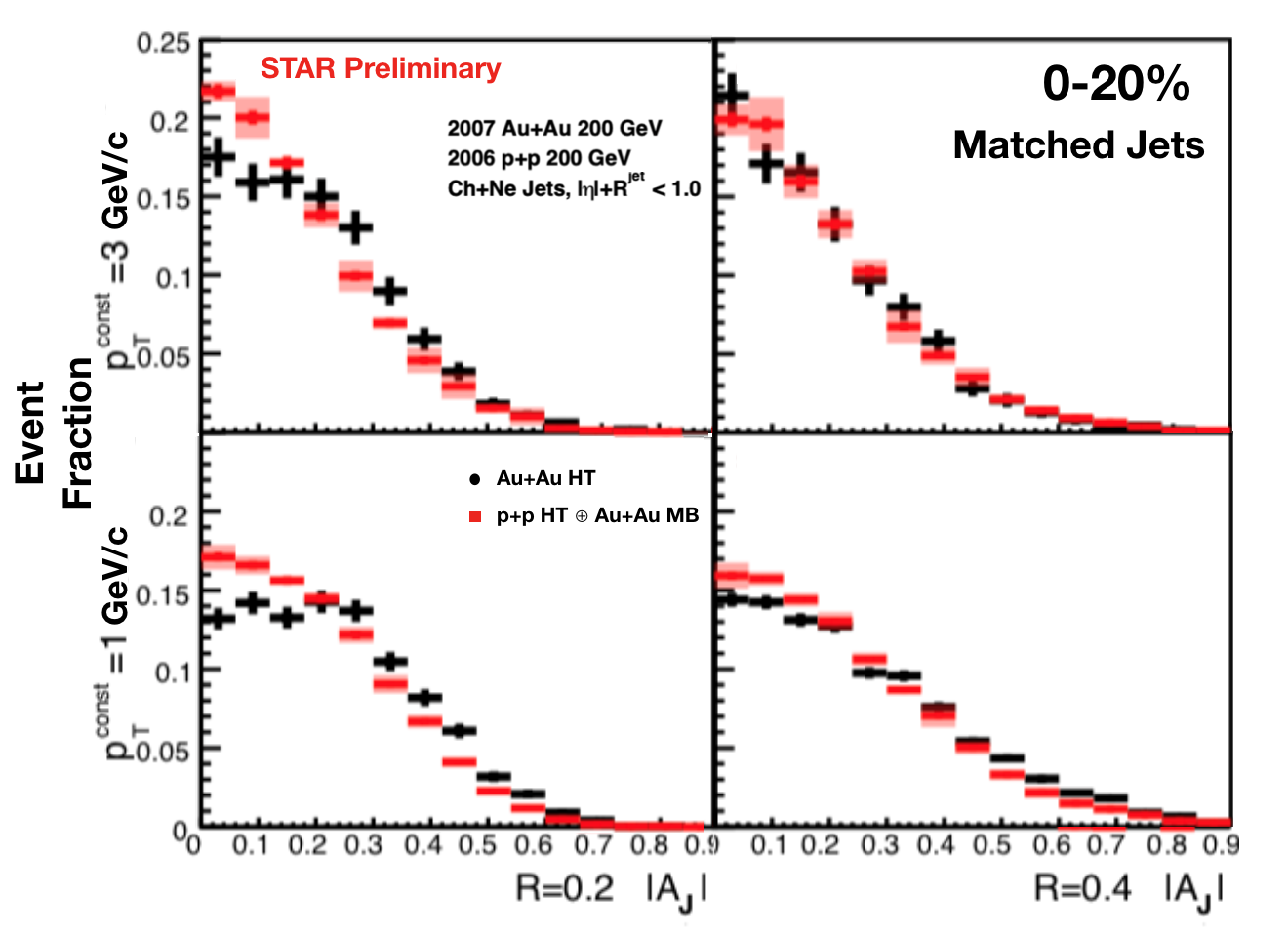}
   \caption{HardCore (left) and Matched (right) di-jet asymmetry ($|A_{\rm{J}}|$) for varying jet finding parameters including jet radius and Hardcore constituent threshold. Markers are described in the text.}
   \label{fig:diffaj}
\end{figure}

Due to the reduced center of mass energy at RHIC than the LHC, the requirement of a high momentum hadron trigger in jet selection biases the jet production vertex towards the periphery of the interaction region~\cite{surfacebias}. The level of such surface bias at RHIC energies can be further controlled by selecting di-jets with asymmetric momentum thresholds on the trigger and its recoil jet, or changing the constituent momentum or energy thresholds. For HardCore jets with a higher constituent threshold, we further bias the trigger jet to be produced on the surface which enforces a larger comparative path length for the recoil jets and thus greater energy loss. In varying the constituent threshold of the HardCore jets and the jet finding radius, we select different population of di-jets that could have different path lengths and thus quantitatively different extents of energy loss. This is the main idea behind a technique introduced by STAR called Jet Geometry Engineering (JGE). With JGE we differentially measure di-jet asymmetry $A_{\rm{J}}$  as shown in Fig.~\ref{fig:diffaj} for both the HardCore (left) and Matched di-jets (right) for jets of radii $0.2$ and $0.4$ (left and right columns, respectively) with the constituent threshold varied from $1.0$ GeV/$c$ and $3$ GeV/$c$ (top and bottom rows respectively). The $A_{\rm{J}}$ results from Au+Au events are shown in the black markers in comparison to the $p$+$p$ reference in the red markers, while the systematic uncertainty on the embedded reference is shown in the red shaded boxes. 

For all values of the constituent threshold and jet radius, we find the HardCore di-jets are imbalanced, with the black and red markers being significantly different from each other. This imbalance is due to energy loss for the recoil jets. In the Matched di-jets on the other hand, they are still imbalanced for the lower constituent threshold and smaller jet radius as shown in the bottom left panel of Figure~\ref{fig:diffaj}. As we increase the jet radius and HardCore constituent thresholds, they become balanced as seen in the red and black markers being indistinguishable within uncertainties. This refers to the lost energy being now recovered within a cone of jet radius of $R=0.4$ and found in the softer constituents included in the Matched jets. From these differential studies, we can conclude based on the assumption of surface bias that the recoil jet with larger jet radius and a higher HardCore constituent threshold undergoes a relatively smaller path length in the medium and consequently is less quenched.  

\section{Jet Angular Scale in Au$+$Au Collisions}

The next step in the differential analysis is to estimate the jet quenching dependence on the medium resolution scale. In our search for a jet observable which is related to the opening angle in jets, we found the groomed jet radii (R$_{\rm{g}}$) to be highly sensitive to the fluctuating underlying event in Au$+$Au collisions and therefore we devised a new observable involving sub-jets of a smaller radius reconstructed within the original jet (see here~\cite{SubJetliliana} for a recent theoretical article demonstrating similar classes of observables).
For our nominal anti-$k_{\rm{T}}$ jets of $R=0.4$, we reconstruct an inclusive set of anti-$k_{\rm{T}}$ sub-jets with $R=0.1$ from the original jet's constituents. An absolute minimum sub-jet $p_{\rm{T}}$ requirement of $2.97$ GeV/$c$ is enforced in central Au$+$Au collisions to reduce sensitivity to the background fluctuations. The two observables related to the momentum and angular scales are then defined as follows:  $z_{\rm{SJ}} = \frac{\rm{min}\it{\left( p^{\rm{SJ1}}_{\rm{T}}, p^{\rm{SJ2}}_{\rm{T}}\right)}}{p^{\rm{SJ1}}_{\rm{T}} + p^{\rm{SJ2}}_{\rm{T}}}$, and $\theta_{\rm{SJ}} = \Delta R (\rm{SJ1, SJ2})$, where $\rm{SJ1, SJ2}$ are the leading and sub-leading sub-jets, respectively. 
We find that these new $z_{\rm{SJ}}$ and $\theta_{\rm{SJ}}$ observables have a diagonal correlation between jets in $p$+$p$ events and the same jets once embedded in a Au$+$Au minimum bias event, which makes them robust to the underlying event.  

\begin{figure}[h] 
   \centering
   \includegraphics[width=0.6\textwidth]{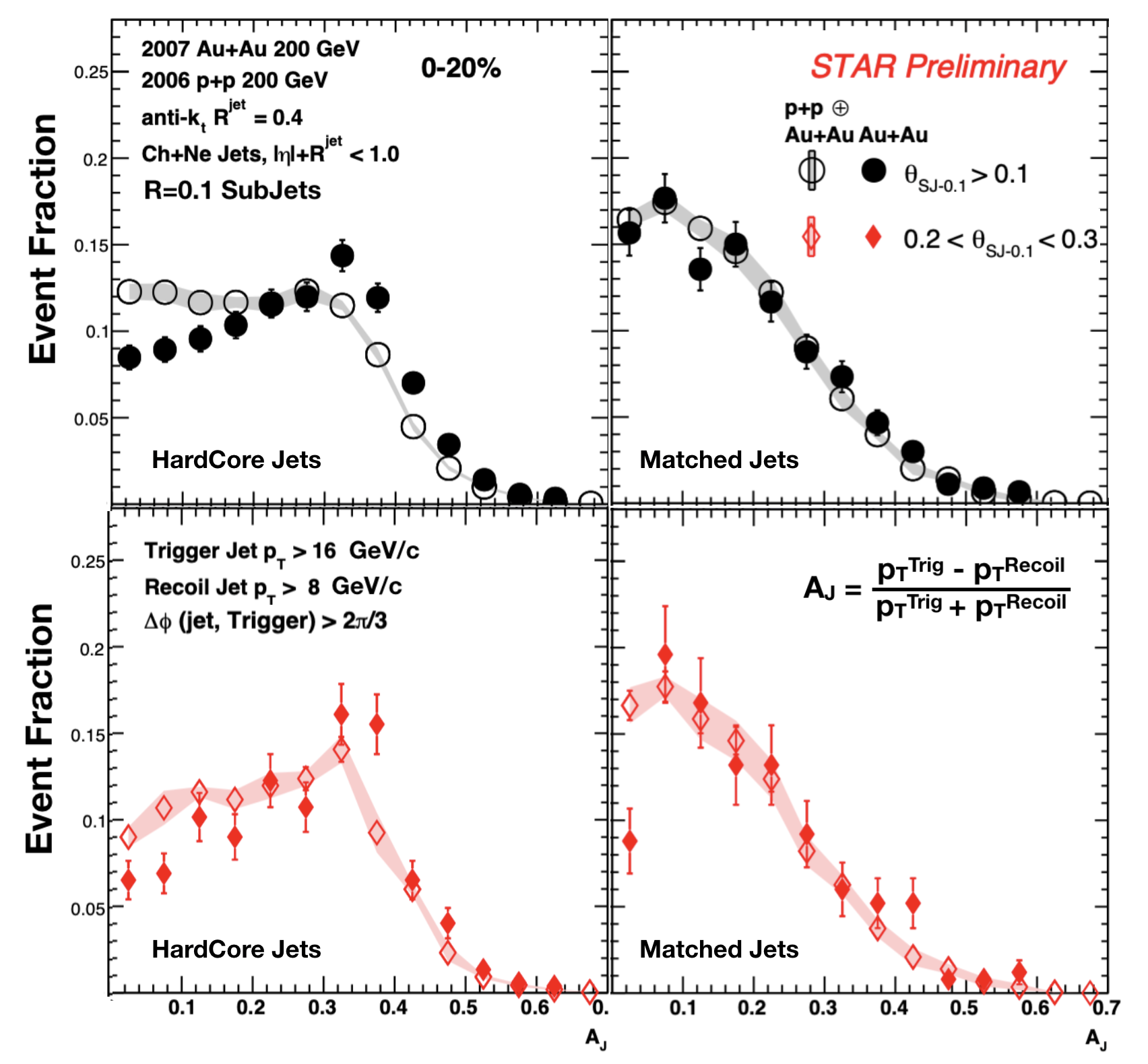} 
   \caption{HardCore and Matched di-jet asymmetry ($|A_{\rm{J}}|$). Markers are described in the text. 
   }
   \label{fig:aj}
\end{figure}

The $z_{\rm{SJ}}$ and $\theta_{\rm{SJ}}$ distributions for Matched jets recoiling off the trigger (selected with a $|\Delta \phi(\rm{jet, HT})| > 2\pi/3$) after constituent subtraction~\cite{cs} are observed to be similar in both Au$+$Au and $p$+$p$ $\oplus$ Au$+$Au. We also observe a remarkable difference in the shape of $z_{\rm{SJ}}$ when compared to that of the SoftDrop $z_{\rm{g}}$, 
which is a consequence of selecting the core of the jet via the leading and subleading subjets. The $\theta_{\rm{SJ}}$ distribution for jets within the considered $p_{\rm{T}}$ range peaks at small values and includes a natural lower cutoff at the sub-jet radius and we now select jets based on this distribution. 

Di-jet asymmetries for both HardCore (left panels) and Matched jets (right panels) are shown in Fig.~\ref{fig:aj}. The black and red markers represent recoil jets with selections on $\theta_{\rm{SJ}}$ $[0.1, 0.4]$ and $[0.2, 0.3]$ for inclusive and wide jets, respectively. We observe a clear di-jet imbalance indicating jet quenching effects in the $|A_{\rm{J}}|$ distributions (comparing Au$+$Au to $p$+$p$ $\oplus$ Au$+$Au) for all HardCore jets including the wide angle jets. The Matched jets on the other hand are balanced at RHIC energies which is consistent with our earlier measurement~\cite{starprl}. We also note that wide angle jets are still balanced indicating no apparent distinction between wide and narrow jets by the medium in our selection. Further detailed differential analyses are required with the high statistics 2014 data set to extract the medium resolution scale or the coherence length and the effect on standard jet quenching observables at RHIC energies.

\section{Conclusions}
STAR has explored the possibility to vary the energy loss for di-jets by modifying the jet finding parameters via JGE. With JGE we can select di-jets that are relatively more or less modified compared to a $p$+$p$ reference. Further studies are planned that may help constrain the path length dependence of partonic energy loss in the QGP at RHIC energies. Due to the sensitivity of the SoftDrop observables to the Au$+$Au underlying event, we introduce the TwoSubJet observables, $z_{\rm{SJ}}$ and $\theta_{\rm{SJ}}$ for $R=0.1$ anti-$k_{\rm{T}}$ sub-jets to represent the momentum and angular scales of a jet in a heavy ion environment. We measure the di-jet momentum asymmetry and find that HardCore di-jets are imbalanced and Matched di-jets are balanced for jets of varying angular scales. We find no significant difference in the quenching phenomenon between wide and narrow jets leading to the conclusion that these jets do not undergo significantly different jet-medium interactions despite different angular scales. 

In combining these two differential measurements together we develop a consistent picture of partonic energy loss at RHIC. Taking into consideration JGE and the small path length for the recoil jet in our specially selected di-jets, and the fact that the recovery of the lost energy for the recoil Matched jets are not dependent on the opening angle of these jets, we can estimate that these jets split and fragment at later times and outside the medium. Given that the HardCore jets do undergo energy loss (the $A_{\rm{J}}$ is still imbalanced in JGE), the energy loss by these jets may have happened in earlier times with the individual hard scattered parton traversing the medium. This will be explored in an upcoming publication with a larger dataset from STAR taken in 2014 wherein we can quantitatively estimate the energy loss due to medium-induced radiation from a single color charge. Following such a measurement we have the opportunity to reduce the di-jet bias and select jets with longer path length and larger opening angles to tease out the resolution scale or coherence length dependence of jet quenching. 

\bibliographystyle{elsarticle-num}







\end{document}